\documentclass[12pt]{article}
\usepackage[utf8]{inputenc}
\usepackage[pdftex]{graphicx}
\usepackage{amsmath,amsthm,amssymb}
\usepackage{mathrsfs}
\usepackage{eucal}
\usepackage{mathtools}
\usepackage{nameref}        
\usepackage[most]{tcolorbox}

\usepackage[unicode=true,bookmarks=true,bookmarksnumbered=false,bookmarksopen=false,breaklinks=false,backref=false,colorlinks=true,urlcolor=black,citecolor=black,linkcolor=black]{hyperref}
\usepackage{xspace}
\usepackage{cancel} 
\usepackage[toc,page]{appendix} 

\usepackage{pgfplots}
\pgfplotsset{compat=1.16}

\usepackage{listings}

\theoremstyle{plain}

\theoremstyle{definition}

\theoremstyle{remark}

\usepackage{xspace}

\usepackage{tikz}
\usepackage{pgfplots}
\pgfplotsset{compat=1.16} 
\tikzstyle{vertex} = [fill,shape=circle,node distance=30pt]
\tikzstyle{edge} = [fill,opacity=.6,fill opacity=.5,line cap=round, line join=round, line width=10pt]
\tikzstyle{elabel} =  [fill,shape=circle,node distance=30pt,fill opacity=.9]
\pgfdeclarelayer{background}
\pgfsetlayers{background,main}
\usepackage[most]{tcolorbox}
\definecolor{mygray}{gray}{0.95}

\usepackage{amsmath}
\usepackage{blkarray}

\usepackage{subfigure}
\usepackage{graphicx}  

\usepackage{framed}

\usepackage{lipsum}
\usetikzlibrary{positioning,chains,fit,shapes,calc}

\usepackage[margin=0.5in]{geometry} 
\usepackage[utf8]{inputenc}
\usepackage{tikz,pgfplots,ulem}
\usepackage{xcolor}
\usetikzlibrary{arrows.meta}
\definecolor{brightcerulean}{rgb}{0.11, 0.67, 0.84}
\usetikzlibrary{positioning,shadows,arrows,trees,shapes,fit}
\usepackage[most]{tcolorbox}

\usepackage{pdflscape}

\usepackage{authblk}

\usepackage{makecell}
\usepackage{setspace}

\title{\textbf{HAT: Hypergraph Analysis Toolbox}}

\author[1,2]{Joshua Pickard}
\author[3]{Can Chen}
\author[4]{Rahmy Salman}
\author[1]{Cooper Stansbury}
\author[4]{Sion Kim}
\author[5]{Amit Surana}
\author[6]{Anthony Bloch}
\author[1,2,6,*]{Indika Rajapakse}

\affil[1]{\small{Department of Computational Medicine and Bioinformatics, University of Michigan, Ann Arbor, MI 48109 USA}}
\affil[2]{iReprogram, Inc., Ann Arbor, MI 48105 USA}
\affil[3]{Channing Division of Network Medicine, Department of Medicine, Brigham and Women’s Hospital and Harvard Medical School, Boston, MA 02115 USA}
\affil[4]{Department of Electrical Engineering and Computer Science, University of Michigan, Ann Arbor, MI 48109 USA}
\affil[5]{Raytheon Technologies Research Center, East Hartford, CT 06108 USA}
\affil[6]{Department of Mathematics, University of Michigan, Ann Arbor, MI 48109 USA}
\affil[*]{\small To whom correspondence should be addressed (indikar@umich.edu).}
\date{}

\begin{document}
\maketitle
\begin{abstract}
\noindent Recent advances in biological technologies, such as multi-way chromosome conformation capture (3C), require development of methods for analysis of multi-way interactions. Hypergraphs are mathematically tractable objects that can be utilized to precisely represent and analyze multi-way interactions. Here we present the Hypergraph Analysis Toolbox (HAT), a software package for visualization and analysis of multi-way interactions in complex systems.
\end{abstract}



\section{Introduction}
Network science is a powerful framework for studying complex systems. However, recent work highlights the limitations of classical methods in networks, which only consider pairwise interactions between nodes to describe group interactions. Use of hypergraphs, in which an edge can connect more than two nodes, has therefore emerged as a new frontier in network science \cite{battiston2020networks,benson2021higher}.

Chromosome conformation capture (3C) methods identify physical interactions (``contacts") between genomic loci \cite{dekker2002capturing, liebermanaiden2009}. While classical capture is pairwise, recent advancements capture multi-way chromatin interactions via proximity ligation (Pore-C, Supplementary Information 5.1) \cite{deshpande2022identifying}, split-pool tagmentation (SPRITE) \cite{quinodozHigherOrderInterchromosomalHubs2018}, or multi-contact 3C (MC-3C) \cite{tavares2020multi}. However, the investigation and biological interpretation of these multi-way contacts is hampered by scarcity of methods for multi-way data \cite{deshpande2022identifying, dotson2022deciphering}.
Hypergraphs are a mathematically tractable extension of graph theory that precisely represent multi-way interactions (Supplementary Information 5.2) \cite{benson2021higher}.
We introduce the Hypergraph Analysis Toolbox (HAT), a general purpose software for the analysis of multi-way interactions and higher-order structures. HAT contains both well-studied and novel mathematical methods for hypergraph analysis in both MATLAB and Python.

Motivated to investigate Pore-C data, HAT is designed as a versatile software for hypergraph analysis.
While there are several robust libraries for graph analysis, most hypergraph software is not multi-faceted and targets specific problems, such as hypergraph partitioning or clustering (Table \ref{tab: HG software}). 
As a general purpose tool, the algorithms implemented in HAT address hypergraph construction, visualization, and the analysis of structural and dynamic properties. HAT is the first software to utilize tensor algebra for hypergraph analysis \cite{chen2020tensor, chen2021controllability, surana2021hypergraph}, and it contains recently developed methods for hypergraph similarity measures \cite{surana2021hypergraph}. HAT is open source, standardized across MATLAB (version 2021b onward) and Python (version 3.7 onward) implementations, and is documented at \url{https://hypergraph-analysis-toolbox.readthedocs.io}, where it will continue to be maintained and developed.

\begin{table}[h]
    \centering
    \begin{tabular}{|l|l|l|}
    \hline
    \multicolumn{1}{|c|}{\textbf{Software}}&\multicolumn{1}{c|}{\textbf{Language}}&\multicolumn{1}{c|}{\textbf{Features}}\\
    \hline
    Hypergraph Analysis Toolbox& MATLAB and Python &\makecell[l]{construction, visualization, expansion, similarity\\ measures, centrality, entropy, tensor based\\ analysis, controllability}\\
    \hline
    HyperNetX \cite{hyperNetX2022software}&Python&clustering, visualization, homology, clustering\\
    \hline
    HALP \cite{halp2014}&Python&directed hypergraphs, walks, partitioning\\
    \hline
    hMETIS \cite{karypis1998hmetis}&C/C++&partitioning, implemented in parallel\\
    \hline
    Phoenix \cite{kurte2021phoenix}&C/C++&clustering, implemented in parallel\\
    \hline
    \end{tabular}
    \caption{Comparison of HAT to well-documented hypergraph libraries. There are several other notable hypergraph software not listed in the table \cite{chgl2019, huang2015scalable, hyperGraphLib2022, multihypergraph2019software}} 
    \label{tab: HG software}
\end{table}


\section{Applications}
\begin{figure}[t!]
    \includegraphics[width=\textwidth]{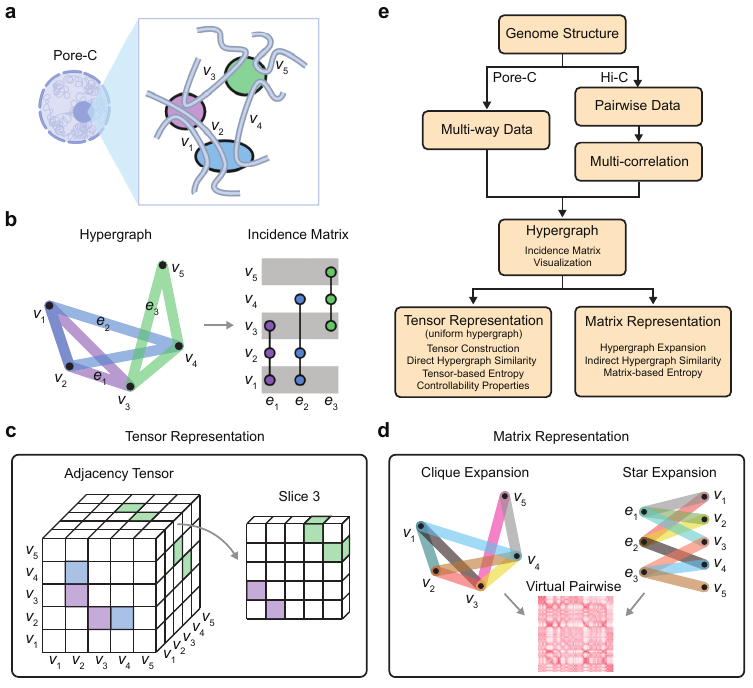}
    \caption{\textbf{a.} The Pore-C assay identifies multi-way chromatin strand colocalization within the nucleus \cite{deshpande2022identifying}. \textbf{b.} Hypergraph representation of Pore-C is drawn where each chromatin strand is represented as a vertex and the multi-way contacts are hyperedges. This is depicted as both a hypergraph and an incidence matrix. \textbf{c.} For multi-way contacts of uniform size, hypergraphs are numerically represented as an adjacency tensor or multi-dimensional matrix. \textbf{d.} Multi-way structure are decomposed with clique and star expansions that generate virtual pairwise contacts \cite{deshpande2022identifying}. \textbf{e.} The workflow of HAT to construct hypergraphs from data, visualize, represent numerically, and computations available for each representation are outlined as a flowchart.}
    \label{fig:1}
\end{figure}

In the work of \cite{dotson2022deciphering}, methods contained in HAT were utilized to examine Pore-C data (Figure \ref{fig:1}a, Supplementary Information 5.1). 
Hypergraphs were constructed from Pore-C data from multiple cell types. Hypergraph entropy measures the structural organization of the genome. Hypergraph similarity measures were utilized to compare the structural similarity between different regions of the genome and cell types. This hypergraph analysis was integrated with other sequencing modalities to identify transcriptional clusters and elucidate the higher-order organization of the genome. Other applications of HAT include investigating social networks \cite{luqman2019complex, arya2018exploiting}, supply chain networks \cite{suo2018exploring, yu2022supply}, (bio)chemical reaction networks \cite{jost2019hypergraph, chen2022teasing}, and epidemiological and ecological networks \cite{battiston2020networks, benson2021higher, golubski2016ecological,bodo2016sis}.






\section{Methods}
HAT can visualize and analyze multi-way interactions. The incidence matrix is the primary representation of hypergraphs in HAT (Figure \ref{fig:1}b) \cite{dotson2022deciphering, beauguittePaohvisOutilPour2020}. HAT targets the following hypergraph features and problems: (1) construction from data \cite{drezner1995multirelation,wang2014measures,taylor2020multi}, (2) expansion and numeric representation \cite{rodriguez2003laplacian,bolla1993spectra,zhou2005beyond}, (3) characteristic structural properties (such as entropy \cite{chen2020tensor}, centrality \cite{tudisco2021node}, distance \cite{surana2021hypergraph}, and clustering coefficients \cite{chen2020tensor}), (3)  controllability \cite{chen2021controllability}, and (4) similarity measures \cite{surana2021hypergraph}. The workflow for using HAT is outlined in Figure \ref{fig:1}e.


\textbf{Construction from Data.}
There are two approaches for constructing a hypergraph from data (Supplementary Information 5.2). Data formats with explicit multi-way interactions, such as Pore-C are directly input to HAT for hypergraph construction. However, the vast majority of data are either pairwise observations (e.g., Hi-C) or do not contain either pairwise or multi-way interactions (e.g. sequencing data), so we implemented three measures to infer multi-way relationships based on multi-correlation measures 
\cite{drezner1995multirelation, wang2014measures,taylor2020multi}. 
HAT constructs hyperedges by setting a minimum threshold for the multi-correlation to be considered a hyperedge. 

\textbf{Expansion and Numerical Representation.}
For uniform hypergraphs, the adjacency, degree, and Laplacian tensors (Figure \ref{fig:1}c) are provided and utilized in similarity, entropy, and controllability calculations (Supplementary Information 5.3). Such tensor based calculations are not currently supported for non-uniform hypergraphs and will be pursued in the future. However, both uniform and non-uniform hypergraphs expand to pairwise structures (Figure \ref{fig:1}d, Supplementary Information 5.4). HAT contains hypergraphs expansions to generate clique expansions, star expansions, and line graphs. These representations facilitate indirect hypergraph similarity and entropy measures for non-uniform hypergraphs. Each hypergraph expansion has unique adjacency, degree, Laplacian, and normalized Laplacian matrices \cite{rodriguez2003laplacian,bolla1993spectra,zhou2005beyond}.

\textbf{Characteristic Structural Properties.}
The following structural properties of hypergraphs are computed: average distance between vertices is computed based on Equation (30) in \cite{surana2021hypergraph}; the clustering coefficient is calculated with Equation (11) in \cite{chen2020tensor}; hypergraph centrality is measured according to methods in \cite{tudisco2021node}. 
For a uniform hypergraph, entropy is computed according to \cite{chen2020tensor}, which defined hypergraph entropy based on the higher-order singular values of the Laplacian tensor. For non-uniform hypergraphs, standard graph entropy measures are applied to the aformentioned hypergraph expansions.

\textbf{Controllability.}
For a uniform hypergraph, the controllability matrix may be computed given the set of input or control nodes \cite{chen2021controllability}. HAT is the first software to analyze controllability properties of hypergraphs.


\textbf{Similarity Measures.}
Hypergraph similarity is measured according to the recent work \cite{surana2021hypergraph}, which distinguishes direct and indirect hypergraph similarity measures. Direct measures utilize tensor representations of uniform hypergraphs; indirect measures utilize graph similarity measures applied to hypergraph expansions. HAT is the first software to implement hypergraph similarity using a tensor representation based on the novel methods in \cite{surana2021hypergraph}. A series of spectral-based measures, as well as Hamming Distance, the Jaccard Index, and centrality measures are provided to measure the similarity between hypergraphs.

For ease of use, the MATLAB and Python implementations are functionally independent but syntactically similar. The software may be installed from the online documentation, GitHub, or via PIP and the MathWorks file exchange for the respective Python and MATLAB implementations.


\section{Conclusion}
Hypergraphs can represent multi-way relationships unambiguously. HAT offers visualization and a computational framework for studying hypergraphs and Pore-C data. Thus HAT can advance the study of multi-way interactions in the genome or other complex systems.





\section*{Acknowledgments}
This work is supported in part by the Air Force Office of Scientific Research (AFOSR) awards FA9550-18-1-0028 and  FA9550-22-1-0215 and 
 NSF DMS2103026 and a  MathWorks Fellowship to the Rajapakse lab.


\bibliographystyle{unsrt}
\bibliography{bibliography}

\section{Supplementary Information}

\subsection{Pore-C: multi-contact, chromosome conformation capture}

Pore-C is a long read sequencing technique designed to capture structural features of genome architecture \cite{deshpande2022identifying,dotson2022deciphering}. It is the most recent extension of chromosome conformation capture (3C) technologies \cite{liebermanaiden2009}. Pore-C data contains multi-way contacts indicating sets of genomic loci that are colocalized in the nucleus. This reveals insight into the higher-order spatial organization of the genome. The Pore-C assay contains similar processes to previous 3C methods \cite{deshpande2022identifying}. First, multi-way contacts between any number of genomic loci are ligated in the nucleus. The genomic loci in these regions are detached from their original chromosomes and chained together. The chained regions are sequenced to determine the set of genomic loci that were originally collocated together. Hypergraphs are natural representations of Pore-C data \cite{dotson2022deciphering}. Individual genomic loci, which can be viewed and binned at any scale for this representation, are represented as vertices in the hypergraph and the colocalization of multiple loci defines a hyperedge. Hi-C data, which captures similar colocalized pairwise relationships, is commonly represented as the adjacency matrix of a graph, but the multi-way contacts of Pore-C necessitate its representation as a hypergraph. The Pore-C assay has already contributed to the field, and new methods of analyzing this data continue to be developed \cite{deshpande2022identifying,dotson2022deciphering}.

\subsection{Hypergraphs}
Hypergraph theory extends graph structures to represent multi-way relationships among elements of a set. Mathematically, a graph $\mathcal{G}=\{\mathcal{V},\mathcal{E}_g\}$ is a set of vertices $\mathcal{V}$ together with a set of edges $\mathcal{E}_g,$ where each edge $e\in \mathcal{E}_g$ is a pair of vertices (i.e., $e=(v_i,v_j)$ where $v_i,v_j\in \mathcal{V}$). Graphs are numerically represented as matrices.

Hyperedges model multi-way relationships by allowing hyperedges to contain any number of vertices, expanding beyond the pairwise restrictions of a graph. Formally, a hypergraph $\mathcal{H}=\{\mathcal{V},\mathcal{E}_h\}$ is a set of vertices together with a set of hyperedges $\mathcal{E}_h$ where each hyperedge $h\in \mathcal{E}_h$ is a subset of vertices (i.e., $h\subseteq \mathcal{V}$). When all hyperedges of a hypergraph have cardinality $k$, it is referred to as a $k$-uniform hypergraph. The extension from edges to hyperedges makes hypergraphs more precise representations of data and presents computational advantages.

\subsection{Numeric Representations of Hypergraphs}
The incidence matrix is the primary numerical representation of hypergraphs (Figure \ref{fig:1}b). An incidence matrix $\textbf{H}$ of a hypergraph $\mathcal{H}=\{\mathcal{V},\mathcal{E}_h\}$ is a $n\times m$ matrix when there are $n$ vertices and $m$ hyperedges. Rows of $\textbf{H}$ are vertices in the hypergraph, and columns are hyperedges. Each element $\textbf{H}_{j,i}$ of the incidence matrix is 1 when  vertex $j$ is a member of or incident to hyperedge $i$ and 0 otherwise.

A $k$-uniform hypergraph can also be represented by a tensor (Figure \ref{fig:1}c). The adjacency tensor of a hypergraph is the higher-order analogue of a graph adjacency matrix. Mathematically, given a $k$-uniform hypergraph $\mathcal{H}=\{\mathcal{V},\mathcal{E}_h\}$  with $n$ vertices, the adjacency tensor is defined as
\begin{equation*}
\textsf{A}\in\mathbb{R}^{\overbrace{n\times\dots\times n}^{\text{$k$ times}}}\text{ where }\textsf{A}_{j_1\dots j_k}=\begin{cases}
\frac{1}{(k-1)!}&\text{if }(j_1,\dots,j_k)\in \mathcal{E}_h\\
0&\text{otherwise}
\end{cases}.
\end{equation*}

Given the adjacency tensor representation, there are analogue degree and Laplacian tensors based on their pairwise definitions \cite{chen2020tensor,chen2021controllability,surana2021hypergraph}.

\subsection{Hypergraph Expansions}

There are two primary pairwise representations of hypergraphs (Figure \ref{fig:1}d). Pairwise representations are often helpful to project multi-way interactions into sets of pairwise interactions or to apply standard graph theoretic operations on hypergraphs.

\textbf{Clique Expansion.} The clique expansion algorithm constructs a graph on the same set of vertices as the hypergraph by defining an edge set where every pair of vertices contained within the same edge in the hypergraph have an edge between them in the graph. Given a hypergraph $\mathcal{H}=\{\mathcal{V},\mathcal{E}_h\}$, then the corresponding clique graph is $\mathcal{C}=\{\mathcal{V},\mathcal{E}_c\}$ where

\begin{equation*}
    \mathcal{E}_c = \{(v_i, v_j) |\ \exists\  e\in \mathcal{E}_h \text{ where }v_i, v_j\in e\}.
\end{equation*}

This is called clique expansion because the vertices contained in each $h\in \mathcal{E}_h$ forms a clique in $\mathcal{C}$. While the map from $\mathcal{H}$ to $\mathcal{C}$ is well-defined, the transformation to a clique graph is a lossy process, so the hypergraph structure of $\mathcal{H}$ cannot be uniquely recovered from the clique graph $\mathcal{C}$ alone \cite{surana2021hypergraph}.

\textbf{Star Expansion.}
The star expansion of $\mathcal{H}=\{\mathcal{V},\mathcal{E}_h\}$ constructs a bipartite graph $\mathcal{S}=\{\mathcal{V}_s,\mathcal{E}_s\}$ by introducing a new set of vertices $\mathcal{V}_s=\mathcal{V}\cup \mathcal{E}_h$ where some vertices represent hyperedges. There exists an edge  between each vertex $v,e\in \mathcal{V}_s$ when $v\in \mathcal{V}$, $e\in \mathcal{E}_h,$ and $v\in e$. Each hyperedge in $\mathcal{E}_h$ induces a star in $\mathcal{S}$. This is a lossless process, so the hypergraph structure of $\mathcal{H}$ is well-defined given a star graph $\mathcal{S}$.







\end{document}